\documentclass[11pt]{article}
\usepackage{graphicx}
\usepackage{color}
\usepackage{amssymb}
\usepackage{makecell}
\begin{document}

\begin{center}
\textbf{\Large Gaussian Process State-Space Modeling and Particle Filtering for Time Series Decomposition and Nonlinear Signal Extraction}

\vspace{7mm}
{\large Genshiro Kitagawa\\[3mm]

Tokyo University of Marine Science and Technology\\
and\\
The Institute of Statistical Mathematics
}

\vspace{3mm}
{\today}
\end{center}

\vspace{2mm}

\begin{center}{\bf\large Abstract}\end{center}

\begin{quote}
Gaussian-process state-space models (GP-SSMs) provide a flexible nonparametric alternative for modeling time-series dynamics that are nonlinear or difficult to specify parametrically. While the Kalman filter is effective for linear-Gaussian trend and seasonal components, many real-world systems require more expressive representations. GP-SSMs address this need by learning transition functions directly from data, while particle filtering enables Bayesian state estimation even when posterior distributions deviate from Gaussianity. This paper develops a particle-filtering framework for GP-SSM inference and compares its performance with the Kalman filter in trend extraction and seasonal adjustment. We further evaluate nonlinear signal-extraction tasks, demonstrating that GP-SSMs can recover latent states under sharp or asymmetric dynamics. The results highlight the utility of combining GP modeling with sequential Monte Carlo methods for complex time-series analysis.
\end{quote}

\vspace{5mm}

\section{Introduction}\label{introduction}

State-space modeling has long served as a fundamental framework for analyzing time series that exhibit trend, seasonality, or latent dynamic structure. In the linear-Gaussian setting, the Kalman filter provides an optimal and computationally efficient method for state estimation and has been widely used for trend extraction and seasonal adjustment in both the natural and social sciences\cite{KG 1996}\cite{DK 2012}\cite{SS 2017}. However, many real-world time series are characterized by nonlinear relationships or non-Gaussian variability, for which the classical Kalman filter becomes inadequate\cite{Kitagawa 1987}. Particle filtering methods offer a flexible alternative, allowing posterior state distributions to be approximated even in highly nonlinear systems\cite{GSS 1993}\cite{Kitagawa 1996}\cite{DFG 2001}. Nevertheless, applying particle filters requires an explicit parametric specification of the state transition function, which can be restrictive or unrealistic in applications where the underlying dynamics are unknown or difficult to model.

Gaussian-process state-space models (GP-SSMs) address this limitation by replacing the parametric transition function with a nonparametric Gaussian process (GP) prior. This formulation enables the state dynamics to be learned directly from data while naturally quantifying the associated uncertainty. GP-SSMs have demonstrated strong performance in nonlinear system identification and have the potential to capture complex structures that are difficult to represent in traditional parametric models\cite{FLSR 2013}\cite{FCR 2014}.

In this paper, we investigate state estimation for GP-SSMs using particle filtering. The combination of Gaussian-process transition modeling and particle-based inference allows us to treat both model uncertainty and nonlinear state evolution in a principled Bayesian manner. To illustrate the utility of this approach, we first examine a trend estimation problem and seasonal adjustment problem, comparing GP-SSM-based estimation with results obtained via the classical Kalman filter. While the Kalman filter yields accurate estimates when the trend and seasonal components are well approximated by linear-Gaussian models, GP-SSMs provide greater flexibility and can accommodate departures from linearity without requiring manual specification of functional forms.

We then consider nonlinear state-space systems where the transition function exhibits strong nonlinearity or asymmetric behavior. Through simulated examples, we demonstrate that GP-SSMs trained on sample trajectories are capable of reconstructing latent states even when the underlying dynamics contain asymmetric transitions or non-smooth behavior. Particle filtering plays a crucial role here, as the posterior state distribution may deviate substantially from Gaussianity, making Kalman-based methods unsuitable.

Overall, our contribution is threefold:
(1) we present a particle-filtering framework for GP-SSM state estimation;
(2) we provide detailed comparative analyses with Kalman-filter-based trend and seasonal adjustment; and
(3) we illustrate the performance of the proposed method in nonlinear signal extraction tasks.
These results highlight the advantages of combining nonparametric Gaussian-process modeling with sequential Monte Carlo inference for analyzing complex time series data.

\section{Gaussian Process State-Space Modeling of Time Series}

\subsection{Ordinary State-Space Model}

Given time series $y_1,\ldots ,y_N$, consider the state-space model
\begin{eqnarray}
x_n &=& Fx_{n-1} + Gv_n \nonumber \\
y_n &=& Hx_n + w_n,
\end{eqnarray}
where $x_n$, $v_n$ and $e_n$ are the $m$-dimensional state vector,
$\ell$-dimensional system noise and $1$-dimensional observation noise, respectively.
It is assumed that the system noise $v_n$ and the observation noise $w_n$
are distributed according to Gaussian white noises;
$v_n\sim N(0,Q)$ and $w_n \sim N(0,\sigma^2 )$, respectively.
Stationary time series models such as AR and ARMA models, as well as non-stationary time series models such as trend models and seasonal adjustment models, can be represented in a unified manner within the framework of state-space models.

Then, given a time series, $y_1,\ldots ,y_N$, the predicted, the filtered and the smoothed estimates of the state, $x_{n|n-1}$, $x_{n|n}$, $x_{n|N}$ and their covariances
$v_{n|n-1}$, $v_{n|n}$ and $v_{n|N}$
are obtained by the Kalman filter and the fixed-interval smoothing algorithm.
If the state-space model contains some unknown parameters $\theta = (\theta_1,\ldots ,\theta_k)^T$
the log-likelihood is defined by 
\begin{eqnarray} 
\ell (\theta ) &=& -\frac{1}{2}\biggl\{ N\log 2\pi + \sum_{n=1}^N \log r_n 
                   + \sum_{n=1}^N \frac{\varepsilon_n^2}{r_n} \biggr\} 
\end{eqnarray}
where $\varepsilon_n$ and $r_n$ are one-step-ahead prediction error and 
its variance obtained by
\begin{eqnarray}
\varepsilon_n &=& y_n - H x_{n|n-1} \nonumber \\
r_n &=& \sigma^2 + H V_{n|n-1}H^T.
\end{eqnarray}
Then the values of the parameters are estimated by the maximum likelihood method and the
goodness of the model is evaluated by the AIC\cite{KG 1996}\cite{Kitagawa 2021}.

\subsection{Gaussian Process Regression Model}
A Gaussian process with mean function 0 and covariance function $k(x,x')$,  denoted ${\mathcal GP}\big(m(x), k(x,x^{\prime})\big)$ is a collection of random variables
$\{f(x)|x\in X\}$ such that for any finite set of input points
$x_1, x_2,\ldots ,x_n$, the corresponding random vector $\left( f(x_1), f(x_2),\ldots ,f(x_n)\right)$ 
follows a multivariate normal distribution with mean vector $(m(x_1), m(x_2), \ldots ,m(x_n))^T$ and covariance matrix $K$ whose $(i,j)$-th element is $k(x_i,x_j)$.
In other words, the Gaussian process defines a probability distribution over functions where the covariance function $k(x,x')$  specifies how function values at different inputs are correlated.
It is also possible to treat $g(x)$ as a Gaussian process; however, for simplicity, this paper assumes that $g(x)$ is a known nonlinear function\cite{RW 2006}\cite{Murphy 2012}.

In the context of machine learning, this covariance function $k(x,x')$ is called kernel function.
In this paper, we use the following kernel functions:

\begin{itemize}
\item Linear kernel: $k(x,x') = x^Tx$
\item Gaussian (rbf) kernel: $k(x,x') = \exp \left( - \frac{(x-x')^2}{\theta}\right) $
\item Exponential kernel: $k(x,x') = \exp \left( - \frac{|x-x'|}{\theta}\right) $
\item Periodic kernel: $k(x,x') = \exp \left\{ \theta_1\cos\left(\frac{(x-x')^2}{\theta_2}\right)\right\} $
\end{itemize}

In a Gaussian process, the kernel function determines the correlation structure between input points and thereby governs the class of functions the model can represent. By choosing an appropriate kernel, one can specify properties such as smoothness, periodicity, and characteristic length scales, which in turn define the prior distribution over functions and the resulting predictive distribution of the Gaussian process.

Moreover, composite kernels-such as linear+rbf or rbf$\times$periodic kernels-enable the modeling of more complex structures, including
\begin{itemize}
\item trend and seasonality,
\item long-term and local variations, and
\item  linear and nonlinear components.
\end{itemize}
Thus, designing a kernel that reflects the characteristics required for the task is a crucial step in Gaussian process modeling.

Given observations $D=\left\{ (x_1,y_1),\ldots ,(x_n,y_n)\right\}$, Gaussian process regression assumes
\begin{eqnarray}
 y_i = f(x_i) + \varepsilon_i,\quad \varepsilon_i \sim N(0,\sigma^2),
\end{eqnarray}
where the latent function $f(x)$ follows a Gaussian process
\begin{eqnarray}
 f(x) \sim GP(m(x),k(x,x')).
\end{eqnarray}
The predictive distribution of $y$ at a new input location $x_*$ can then be derived as follows.
Define the cross-covariance vector between the training inputs and the predictive point, the prior variance at the predictive point, and the noise-augmented covariance matrix by
\begin{eqnarray}
  k_* &=& (k(x_1,x_*),\ldots ,k(x_n,x_*)), \nonumber \\
  k_{**} &=& k(x_*,k_*) \\
  K_y &=& K + \sigma^2 I. \nonumber
\end{eqnarray}
Under the Gaussian process prior, the joint distribution $y=(y_1,\ldots ,y_n)^T$ and $y_*$ is given by 
\begin{eqnarray}
  \left[ \begin{array}{c} y \\ y_* \end{array}\right] \sim
  N\biggl( 0,  \left[ \begin{array}{cc} K_y & k_* \\
                                k_*^T & k_{**} \end{array}\right] \biggl).
\end{eqnarray}
Then given the training data $D$, the predictive distribution of $y_*$ is given by
\begin{eqnarray}
  p(y_*|x_*,D) = N\Bigl(k_*^TK_y^{-1}y, k_{**}-k_*^TK_y^{-1}k_*\Bigr).
\end{eqnarray}

\subsection{Gaussian Process State-Space Model}
A Gaussian process state-space model (GP-SSM) is a nonlinear state-space model in which the unknown transition dynamics are represented by a Gaussian process\cite{FLSR 2013}\cite{FCR 2014}.
Formally, the system is defined as
\begin{eqnarray}
  \begin{array}{ccccccc}
     x_n &=& f(x_{n-1}) + v_n, & {}\quad & v_n &\sim& N(0,\tau^2) \\
     y_n &=& g(x_n) + w_n, & {}\quad & w_n &\sim& N(0,\sigma^2) ,
  \end{array}
\end{eqnarray}
where $x_n$ is the latent (unobserved) state, $y_n$ is the observed time series, and 
$v_n$ and $w_n$ denote process and observation noise, respectively.
In a GP-SSM, the transition function $f(\cdot )$ is modeled as a Gaussian process,
\begin{eqnarray}
   f(x) \sim GP(m(x),k(x,x')),
\end{eqnarray}
which provides a nonparametric prior over possible state transition functions.
This formulation allows the dynamics to be learned directly from data without specifying an explicit parametric form.

Given observations $y_1,\ldots ,y_n$, inference in a GP-SSM typically involves estimating the posterior distribution of the latent states $x_n$ and the hyperparameters of the Gaussian process.
Because the resulting model is nonlinear and non-Gaussian, approximate inference methods such as particle filtering, extended/unscented Kalman filtering, or variational approaches are generally employed.

\subsection{Particle Filter for GP-SSM}
In this subsection, we examine state estimation for the GP-SSM using the sequential Monte Carlo (SMC) methods. Particle filtering provides a flexible framework for approximating the posterior distribution of the latent states\cite{FLSR 2013}.
Consider a GP-SSM of the form
\begin{eqnarray}
  \begin{array}{ccccccc}
     x_n &=& f(x_{n-1}) + v_n, & {}\quad & v_n &\sim& N(0,\tau^2) \\
     y_n &=& g(x_n) + w_n, & {}\quad & w_n &\sim& N(0,\sigma^2) ,
  \end{array}
\end{eqnarray}
where the unknown transition function $f(x)$ is endowed with a Gaussian process prior. 
Given observations $y_1,\ldots ,y_n$, the goal is to estimate the posterior distribution 
$p(x_n|y_{1:n})$.

Because the predictive distribution of
\begin{eqnarray}
  p(x_n|x_{n-1},D)
\end{eqnarray}
obtained from the GP is non-Gaussian and often lacks a closed form, the filtering distribution cannot be propagated analytically. 
Particle filter approximates the posterior using samples $\{x_n^{(1)},\ldots ,x_n^{(m)}\}$.

\noindent\textbf{Prediction Step}

For each particle $x_{n-1}^{(i)}$, draw a sample from the GP-based transition prior
\begin{eqnarray}
   \hat{x}_n^{(i)} \sim p(x_n|x_{n-1}^{(i)},D) = N\left(\mu_f(x_{n-1}^{(i)}),\sigma_f^2(x_{n-1}^{(i)})+\tau^2\right),
\end{eqnarray}
where $\mu_f$ and $\sigma_f^2$ are the GP posterior mean and variance.

\noindent\textbf{Filtering Step}

For each particle is evaluate the likelihood of the observation:
\begin{eqnarray}
   w_n^{(i)} \propto p(y_n|\hat{x}_n^{(i)}) = N\left(y_n;g(\hat{x}_n^{(i)}),\sigma^2\right).
\end{eqnarray}
and normalize the weights so that $\sum_{i=1}^N \hat{w}_n^{(i)} = 1$.
Then obtain $m$ samples $\{x_n^{(1)},\ldots ,x_n^{(m)}\}$ by the resampling, namely by
resample the particles $\hat{x}_n^{(1)},\ldots ,\hat{x}_n^{(m)}$ with replacement 
according to the sampling probabilities $\hat{w}_n^{(1)},\ldots ,\hat{w}_n^{(m)}$.
The resulting  particle set provides a Monte Carlo approximation of the filtering distribution
$ p(x_n|y_{1:n})$.

\noindent\textbf{Log-likelihood}
In the particle filter, the log-likelihood of the GP-SSM is approximated by
\begin{eqnarray}
 \ell (\theta) = \sum_{n=1}^N \log p(y_n|y_{1:N})
   \approx \sum_{n=1}^N \log \biggl( \sum_{j=1}^m w_n^{(j)} \biggr) - N \log m .
\end{eqnarray}

\subsection{Constructing Training Data When States Are Not Directly Observed}
When the latent states of the system are not directly observable, training data for a GP-SSM cannot be obtained in the usual form of paired samples $(x_{n-1},x_n)$. 
Instead, the state trajectories must first be inferred from the available measurements. A common approach is to apply a state estimation method such as a Kalman filter, smoother or a particle smoothing technique to obtain posterior estimates of the latent states. The resulting smoothed state means (or samples from the posterior distribution) are then used as surrogate ^^ ^^ ground truth" states. By pairing the estimated states $\hat{x}_{n-1}$ with their one-step-ahead counterparts $\hat{x}_n$, an approximate training set
$\{(\hat{x}_{n-1},\hat{x}_n)\}_{n=2}^N$ can be constructed. This derived dataset is subsequently used to train the Gaussian process model for the transition function.

If the measurement model includes exogenous inputs, or when the uncertainty of the state estimates is important, the method can be extended by incorporating the posterior variances or by drawing multiple state trajectories, allowing the GP to be trained on a richer representation of the latent dynamics.

\section{Examples}
This section presents four illustrative applications of GP-SSMs. We begin with trend estimation and seasonal adjustment, both solvable by Kalman filtering, to provide familiar benchmarks. We then demonstrate two nonlinear cases: one with asymmetric state dynamics and another involving strong nonlinearity combined with external inputs.

\subsection{Trend Estimation}

\begin{figure}[h]
\begin{center}
\includegraphics[width=70mm,angle=0,clip=]{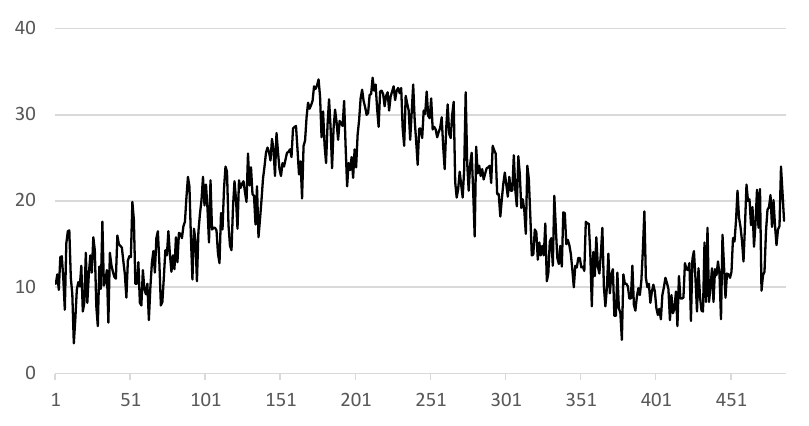}
\caption{Daily maximum temperature data, Tokyo.}
\label{Fig_maxtemp_data}
\end{center}
\end{figure}

In this subsection, we present the results of applying the GP-SSM (Gaussian Process State-Space Model) to the daily maximum temperature data in Tokyo, as shown in Figure \ref{Fig_maxtemp_data}\cite{KG 1996}\cite{Kitagawa 2021}. For comparison, Table \ref{Tab_log-LK_maxtemp} also reports the results obtained using first- and second-order trend models.
\begin{eqnarray}
   \begin{array}{lcl}
      t_n &=& t_{n-1} + v_n \\
      y_n &=& x_n + w_n
   \end{array} \qquad\qquad
   \begin{array}{lcl}
      \left[ \begin{array}{c} t_n \\ \Delta t_n \end{array} \right] &=&
      \left[ \begin{array}{cc} 1 & 1 \\ 0 & 1 \end{array} \right]
      \left[ \begin{array}{c} t_{n-1} \\ \Delta t_{n-1} \end{array} \right] + 
      \left[ \begin{array}{c} 1 \\ 1 \end{array}\right] v_n \\
      y_n &=& x_n + w_n
   \end{array} 
\end{eqnarray}
 Corresponding to the first-order trend model, the following Gaussian process state-space (GP-SSM) models were estimated. 
\begin{eqnarray}
      t_n &=& f(t_{n-1}) + v_n \nonumber\\
      y_n &=& x_n + w_n  \\
     f(x) &=& {\mathcal GP}(0,k(x,x')), \nonumber
\end{eqnarray}
where $k(x,x')$ is the kernel function and $v_n \sim N(0,\tau^2)$ and $w_n \sim N(0,\sigma^2)$. 
For the current example, we compared three kernels: a linear kernel, an rbf (Gaussian) kernel, and a combined linear + rbf kernel. The training data were obtained using a second-order trend model and the Kalman filter. Based on the log-likelihood values, the linear kernel yielded the best performance. However, the log-likelihood of the combined linear + rbf kernel was not substantially different.

\begin{table}[h]
\caption{Log likelihoods of various models for daily maximum temperature data.}
\label{Tab_log-LK_maxtemp}
\begin{center}
\begin{tabular}{l|c|ccccc}
$m$  & log-likelihood & $\sigma^2$& $\tau^2$  & \\
\hline
Kalman ($m=1$)  & $-1224.275$ & 5.544 & 0.224  \\
Kalman ($m$=2)  & $-1255.553$ & 8.079 & $0.322\!\times \!\!10^{-3}$ &    &      \\
\Xhline{1pt}
                  &log-likelihood & $\sigma^2$& $\tau^2$  & $\theta_1$  & $\theta_2$ &$\theta_3$ \\
\hline
GP-SSM(linear)    & $-1220.150$ & 5.574 & $0.258\!\times \!\!10^{-1}$ &      & 1.235 \\
GP-SSM(rbf)       & $-1225.179$ & 0.360 & 3.950 & 0.600 & 1.356 \\
GP-SSM(linear-rbf)& $-1220.595$ & 5.800 & 0.210 & 0.520 & 0.026 \\
2D-GP-SSM(linear) & $-1243.133$ & 8.500 & $0.136\!\times \!\!10^{-2}$  &   & ---  \\
\hline
\end{tabular}
\end{center}
\end{table}

\begin{figure}[tbp]
\begin{center}
\includegraphics[width=100mm,angle=0,clip=]{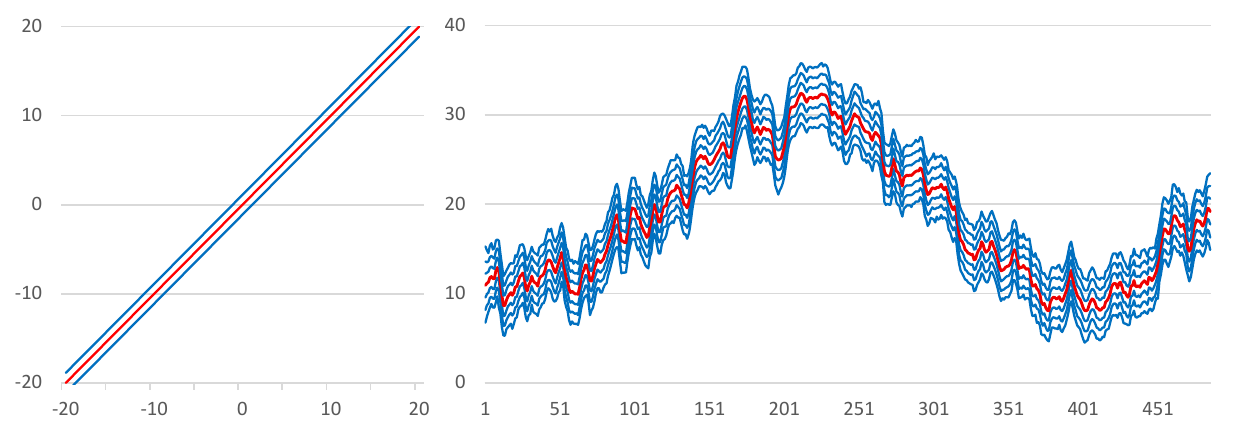}
\caption{Gaussian process regression $f(x)$ and trend of maximum temperature data estimated by 1-D linear karnel}
\label{Fig_maxtemp_linear-kernel}
\includegraphics[width=76mm,angle=0,clip=]{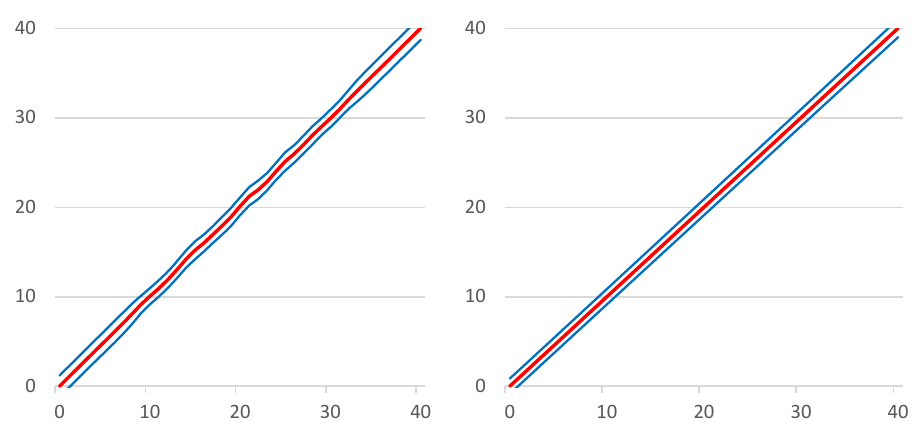}
\includegraphics[width=65mm,angle=0,clip=]{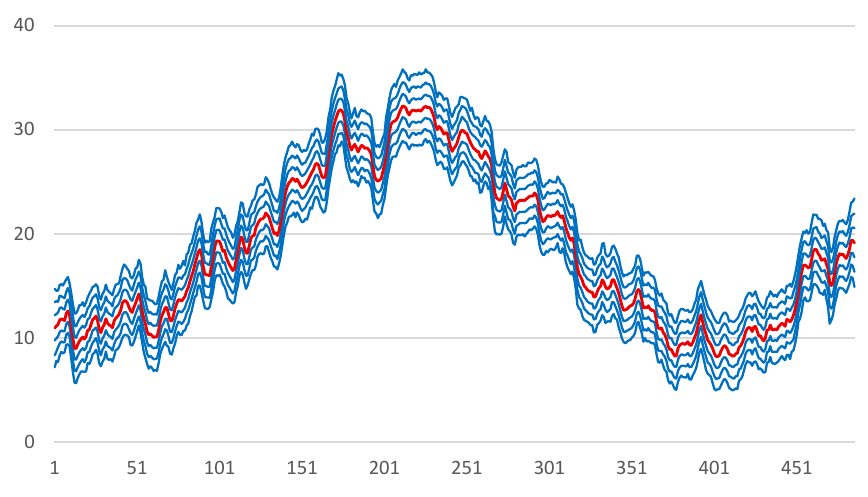}
\caption{1-D linear-rbf karnel: from left to right, the Gaussian process estimated by linear+rbf kernel, linear kernel and estimated trend}
\label{Fig_maxtemp_linear-rbf-kernel}
%
\includegraphics[width=38mm,angle=0,clip=]{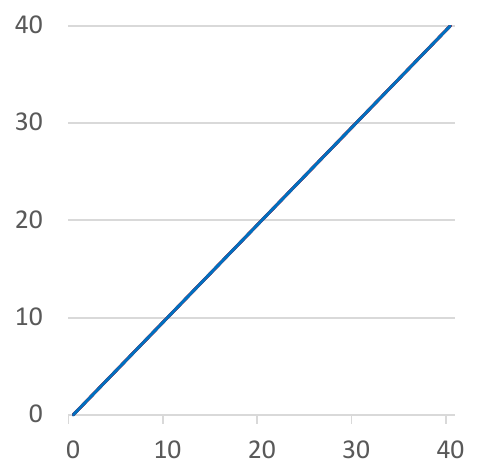}
\includegraphics[width=65mm,angle=0,clip=]{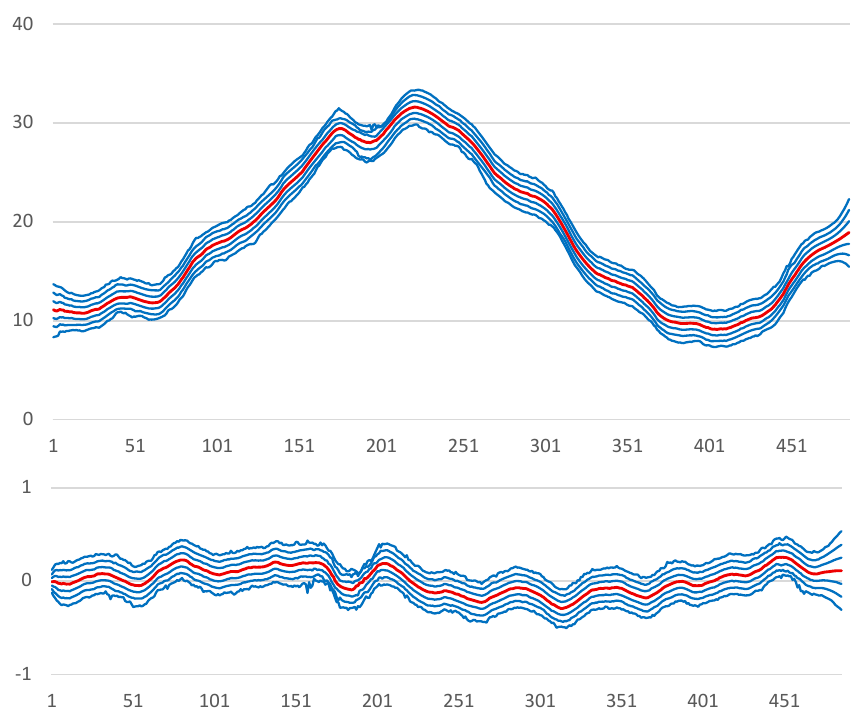}
\caption{Trend of maximum temperature data estimated by 2-D linear and rbf kernels}
\label{Fig_maxtemp_2D-kernel}
\end{center}
\end{figure}

Figure \ref{Fig_maxtemp_linear-kernel} shows the Gaussian process $f(x)$ obtained by using the linear kernel, along with the mean and the $\pm$1, 2, and 3-standard deviation intervals of the estimated trend distribution. As expected, in this case, the estimated values are comparable to the results obtained using the Kalman filter under the assumption that the linear function is known.

\setlength{\thickmuskip}{4mu}
\begin{eqnarray}
x_n &=& x_{n-1} +  v_n \\
y_n &=& x_n + w_n
\end{eqnarray}
where $y_n$, $x_n$, $v_n$ and $w_n$ are the time series, the unobservable trend component, 
the system noise and the observation noise, respectively.
$v_n$ and $w_n$ are assumed to follow the Gaussian white noises such as
$v_n \sim N(0,\tau^2)$ and $w_n \sim N(0,\sigma^2)$.

Figure \ref{Fig_maxtemp_linear-rbf-kernel} presents the Gaussian process obtained with the linear + rbf kernel, the transition function estimated with the linear kernel, and the estimated trend. Although the Gaussian process obtained using the linear + rbf kernel exhibits slight fluctuations, the linear-component part is almost identical to the result shown in Figure \ref{Fig_maxtemp_linear-kernel}, and consequently, the resulting trend is nearly the same as that shown in Figure \ref{Fig_maxtemp_linear-kernel}.

Figure \ref{Fig_maxtemp_2D-kernel} presents the results corresponding to the two-dimensional trend model, where the linear+rbf kernel function $k(x,x')$ is employed. The left panel of Figure \ref{Fig_maxtemp_2D-kernel} shows the Gaussian proces for the trend component, while the upper-right panel shows the estimated trend and the lower-right panel illustrates the differenced trend series. Compared with the results in Figures \ref{Fig_maxtemp_linear-kernel} and \ref{Fig_maxtemp_linear-rbf-kernel}, the obtained trend appears smoother.

\newpage
\subsection{Seasonal Adjustment}
In the standard Decomp model, the time series with seasonality is expressed as
\begin{eqnarray}
y_n &=& T_n + S_n + w_n,
\end{eqnarray}
where $y_n$ is the time series and $T_n$, $S_n$ and $w_n$ are
the trend, the seasonal component and the observation noise, respectively 
\cite{KG 1984},\cite{Kitagawa 2021}.
These components are assumed to follow the component models
\begin{eqnarray}
\Delta^{m_1}T_n &=& v_n^{(T)} \nonumber \\
\left( \sum_{j=0}^{p-1} B^j \right)^{m_2}S_n &=&  v_n^{(S)} \label{eq_component models},
\nonumber
\end{eqnarray}
where $B$ is the back-shift operator defined by $B x_n \equiv x_{n-1}$ and $\Delta = 1-B$
and $w_n$, $v_n^{(T)}$ and $v_n^{(S)}$ are independ Gaussian random variable
with mean 0 and the variances $\sigma^2$, $\tau_1^2$ and $\tau_2^2$, respectively.

The basic observation model (1) and the component models (2) can be expressed in
a state-space model form
\begin{eqnarray}
x_n &=& Fx_{n-1} + Gv_n \nonumber \\
y_n &=& Hx_n + w_n,
\end{eqnarray}
where, for example, for $m_1=2$, $m_2=1$ and $m_3=3$, the state and the system noise are defined by 
$x_n =(T_n, T_{n-1},S_n,$ $S_{n-1},\ldots , S_{n-p+1},p_n, p_{n-1} ,p_{n-2})^T$,
and $v_n =(v_n^{(T)},v_n^{(S)},v_n^{(p)})$, and
$F$, $G$ and $H$ are defined by
{\setlength\arraycolsep{1mm}
\begin{eqnarray}
F &=& \left[ \begin{array}{cc|cccc}\arraycolsep=0.1mm
             2 & -1&   &   &   &       \\
             1 & 0 &   &   &   &       \\ \hline
               &   & -1& -1&\cdots&-1   \\
               &   & 1 &   &   &       \\
               &   &   &\ddots&&       \\
               &   &   &   & 1 &       \\    
       \end{array}\right],\quad
G = \left[ \begin{array}{c|c}
             1 &     \\
             0 &     \\ \hline
               & 1   \\
               & 0   \\
               &\vdots   \\
               & 0   \\ 
       \end{array}\right], \\
H &=& \left[ \begin{array}{cc|cccc} 
                    1 & 0 & 1 & 0 & \cdots & 0  \end{array}\right].
\end{eqnarray}
}
It is also assumed that the system noise $v_n$ and the observation noise $w_n$
are distributed according to Gaussian white noises
$v_n\sim N(0,Q)$ and $w_n \sim N(0,\sigma^2 )$, respectively.

Then, given a time series, $y_1,\ldots ,y_N$, the smoothed estimates of the components, 
$T_{n|N}$ and $S_{n|N}$ are obtained by the Kalman filter and the fixed-interval smoothing algorithm.
The parameters of the model $\theta = (\tau_1^2,\tau_2^2)^T$ is obtained by the maximum likelihood method and the goodness of the model is evaluated by the AIC\cite{Kitagawa 2021}.
\begin{eqnarray}
\hat{\sigma}^2 = \frac{1}{N}\sum_{n=1}^N (y_n - Hx_{n|n-1})^2.
\end{eqnarray}
Then the log-likelihood and the AIC are defined by 
\begin{eqnarray} 
\ell (\theta ) &=& -\frac{1}{2}\left\{ N\log 2\pi + \sum_{n=1}^N \log r_n + N \right\} \nonumber \\
{\rm AIC} &=& -2\ell (\hat{\theta}) + 2(\mbox{number of parameters of the model}),
\end{eqnarray}
where $\varepsilon_n$ and $r_n$ are one-step-ahead prediction error and 
its variance obtained by
\begin{eqnarray}
\varepsilon_n &=& y_n - H x_{n|n-1} \nonumber \\
r_n &=& \sigma^2 + H V_{n|n-1}H^T.
\end{eqnarray}

\subsubsection{IIP Data}
We examine the monthly Index of Industrial Production of Japan from January 1979 to December 2009 (based on the 2020 standard). The number of observations is $N=$360.

Table \ref{Tab_log-LK_asymmetric_AR} summarizes the log-likelihood and AIC values obtained by fitting Decomp models to the IIP data. The trend order ($m_1$) was set to 1 and 2.

\begin{table}[tbp]
\caption{Log likelihoods of various models for IIP data}
\label{Tab_log-LK_asymmetric_AR}
\begin{center}
\begin{tabular}{l|c|ccccc}
$m$  & log-kikelihood & $\sigma^2$ & $\tau_t^2$ & $\tau_s^2$ & $\theta_1$ & $\theta_2$ \\
\hline
Kalman($m_1$=1,$m_3=0$) & 1201.134 & 0.190$\times 10^{-4}$ & 0.9999 &  0.150  &      \\
Kalman($m_1$=2,$m_3=0$) & 1207.778 & 0.206$\times 10^{-4}$ & 0.05608 &  0.205  &      \\
\hline
GP-SSM(true)            & 1030.281 & 0.120$\times 10^{-3}$ & 0.165$\times 10^{-10}$ & 0.150$\times 10^{-2}$ & 0.630 & 0.179 \\
\hline
\end{tabular}
\end{center}
\end{table}

\begin{figure}[tbp]
\begin{center}
\includegraphics[width=70mm,angle=0,clip=]{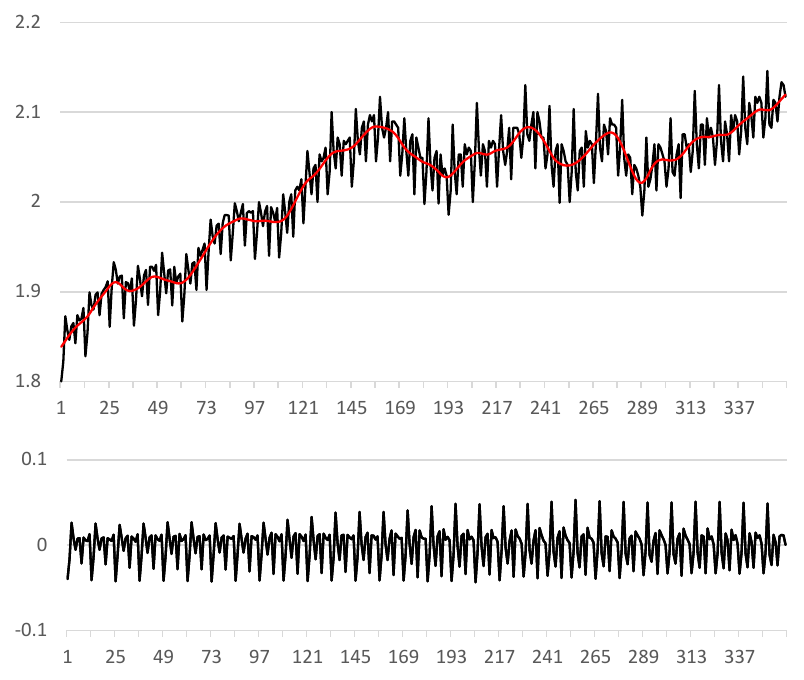}
\caption{IIP data after taking logarithm and detrended and trend and seasonal component estimated by Decomp.}
\end{center}
\end{figure}

In the GP-SSM framework, a two-dimensional linear kernel is used for the trend component, and a periodic kernel is employed for the seasonal component. 
To prevent bias leakage into the seasonal component, the mean of the observed data $y(n)$ is removed prior to modeling. 
The parameters estimated by maximum likelihood are $\sigma^2 = r=0.12\times 10^{-3}$, $\tau_1^2 = 0.165\times 10^{-10}$,
$\tau_2^2 = 0.150\times 10^{-2}$, $\theta_1=0.63$ and $\theta_2=0.179$.

\begin{figure}[tbp]
\begin{center}
\includegraphics[width=100mm,angle=0,clip=]{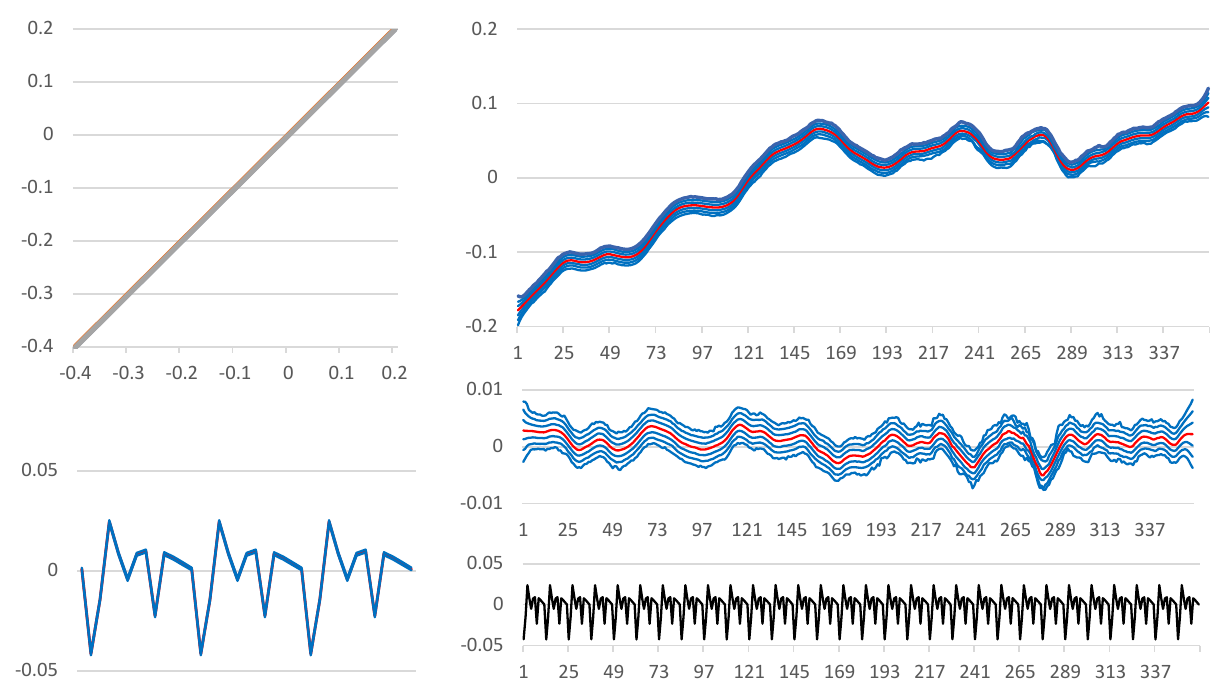}
\caption{IIP data after taking logarithm and detrended.}
\label{IIP-data_GP-SSM}
\end{center}
\end{figure}

Figure \ref{IIP-data_GP-SSM} presents the estimated Gaussian processes corresponding to the trend and seasonal components, along with their derived estimates: the trend, the first-difference of the trend, and the seasonal component. The trend estimate is nearly identical to the result obtained using the Kalman-filter-based state-space model; however, the seasonal component differs in that its fluctuation pattern remains fixed rather than varying over time as in the SSM.

\subsection{Nonlinear AR Model}
In this subsection, we consider the following nonlinear model with observation noise
\begin{eqnarray}
  x_n &=& f(x_{n-1}) + v_n \nonumber \\
  y_n &=& x_n + w_n,
\end{eqnarray}
where $f(x)$ is a nonlinear function defined by
\begin{eqnarray}
f(x) = \left\{ \begin{array}{ll}
           \displaystyle\frac{2b_1x}{x^2 + c^2_1} & x < 0 \\[4mm]
           \displaystyle\frac{2b_2x}{x^2 + c^2_2} & x \geq 0 \end{array}
       \right.
\end{eqnarray}
and $v_n$ and $w_n$ are Gaussian white noises, $v_n \sim N(0,\tau^2)$ , $w_n \sim N(0,\sigma) $.

\begin{figure}[tbp]
\includegraphics[width=50mm,angle=0,clip=]{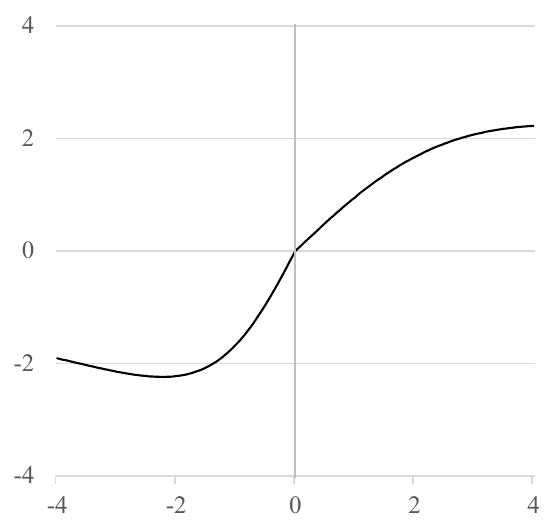}
\hspace{8mm}
\includegraphics[width=100mm,angle=0,clip=]{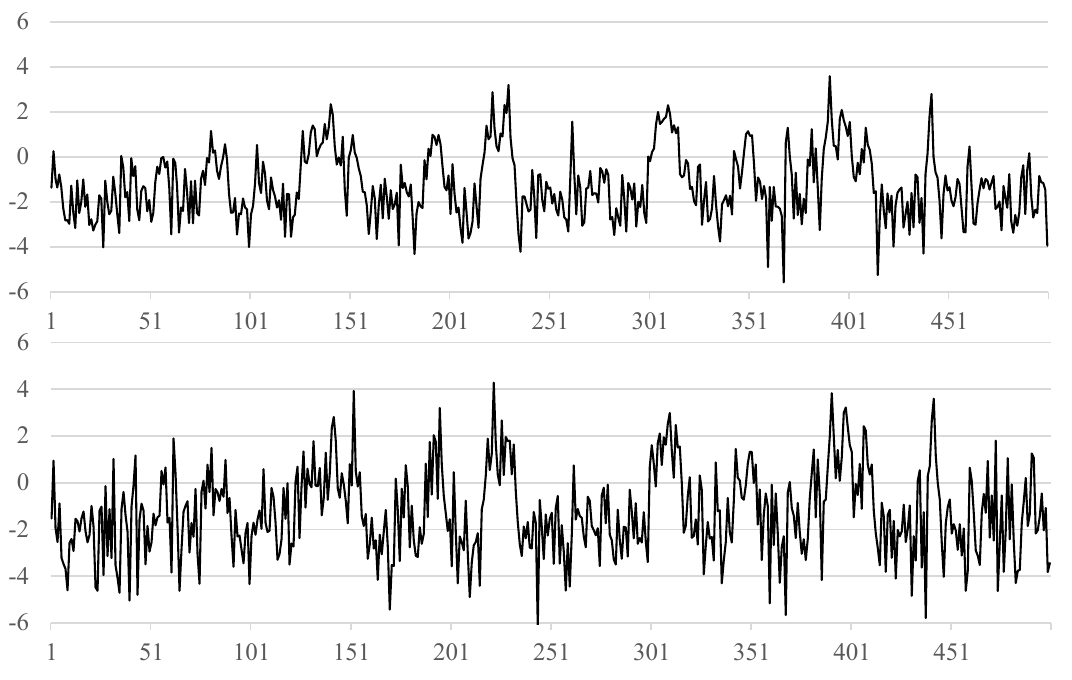}
\caption{Asymmetric function $f(x)$ obtained by $b_1=5$, $c^2_1=5$, $b_2=10$ and $c^2_2=20$ and
test data $x_n$ and $y_n$ generated by $\tau^2 = 1$ and $\sigma^2 = 1$.}
\label{Fig_nonlinear_test-data}
\end{figure}

The left panel of Figure \ref{Fig_nonlinear_test-data} shows the function $f(x)$ for the case where 
$b_1=1$, $c_1^2=5$, $b_2=10$ and $c_2^2=20$.
The top-right and the bottom-right panels of Figure \ref{Fig_nonlinear_test-data} present the generated sequences $x_n$ and $y_n$ assuming $\tau^2=1$ and $\sigma^2=1$, respectively.
Due to the nonlinearity of the function $f(x)$, the generated series $x_n$ fluctuates around approximately 
$-2$ over a long period, occasionally jumping to the positive region.


\begin{figure}[h]
\begin{center}
\includegraphics[width=100mm,angle=0,clip=]{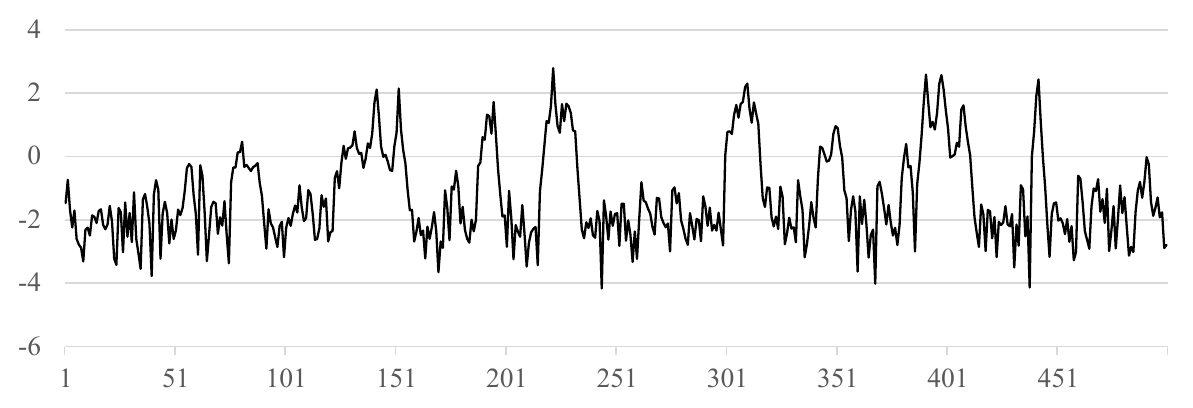}
\caption{Estimated signal by particle filter using the true model.}
\label{Fig_AP-data_particle-filter}
\end{center}
\end{figure}

Figure \ref{Fig_AP-data_particle-filter} shows the results of estimating the latent signal $x_n$ from the time series $y(n)$ using a particle filter, assuming that the parameters of this nonlinear model are known. Although the detailed waveform of the signal is not fully reconstructed, the existence of a stable equilibrium point around $-2$ is reproduced, and the overall shape of the waveform is reasonably well captured. As shown in Table \ref{Tab_log-LK_asymmetric_AR}, when the number of particles is increased to 1,000, $10^4$ and $10^5$, the log-likelihood increases slightly.

\begin{table}[tbp]
\caption{Log-likelihoods and estimated parameters of various models for asymetric AR model}
\label{Tab_log-LK_asymmetric_AR}
\begin{center}
\begin{tabular}{l|cc|c|cc}
               & $m$ &  &log-likelihood & $\tau^2$ & $\sigma^2$  \\
\hline
               & 1,000   &  & -912.065 & 1.0   & 1.0       \\
particle filter& 10,000  &  & -912.046 & 1.0   & 1.0      \\
               & 100,000 &  & -911.985 & 1.0   & 1.0   \\
\hline
AR             &    &             & -941.788 & 0.406 & 1.609  \\
Asymmetric AR  &    &             & -939.477 & 0.400 & 1.616  \\
\hline
\end{tabular}
\end{center}
\end{table}

Table \ref{Tab_log-LK_asymmetric_AR} also presents the estimation results obtained using an AR model ($a=0.939$) with observation noise and an asymmetric AR model. In the asymmetric AR model, the regression coefficients differ depending on the sign of the time series. As shown in Table 3, the maximum likelihood estimates of the coefficients were $a_1=0.868$ and $a_2=0.948$.
As expected, the log-likelihood of this model is approximately 30 lower than that obtained when the true model is assumed to be known.

\begin{figure}[tbp]
\begin{center}
\includegraphics[width=50mm,angle=0,clip=]{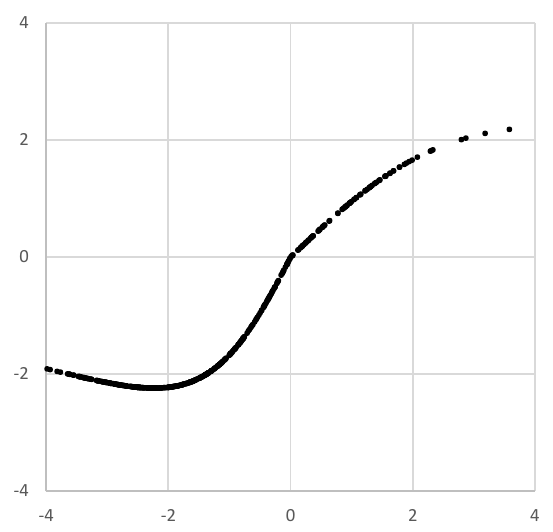}
\includegraphics[width=100mm,height=48mm,angle=0,clip=]{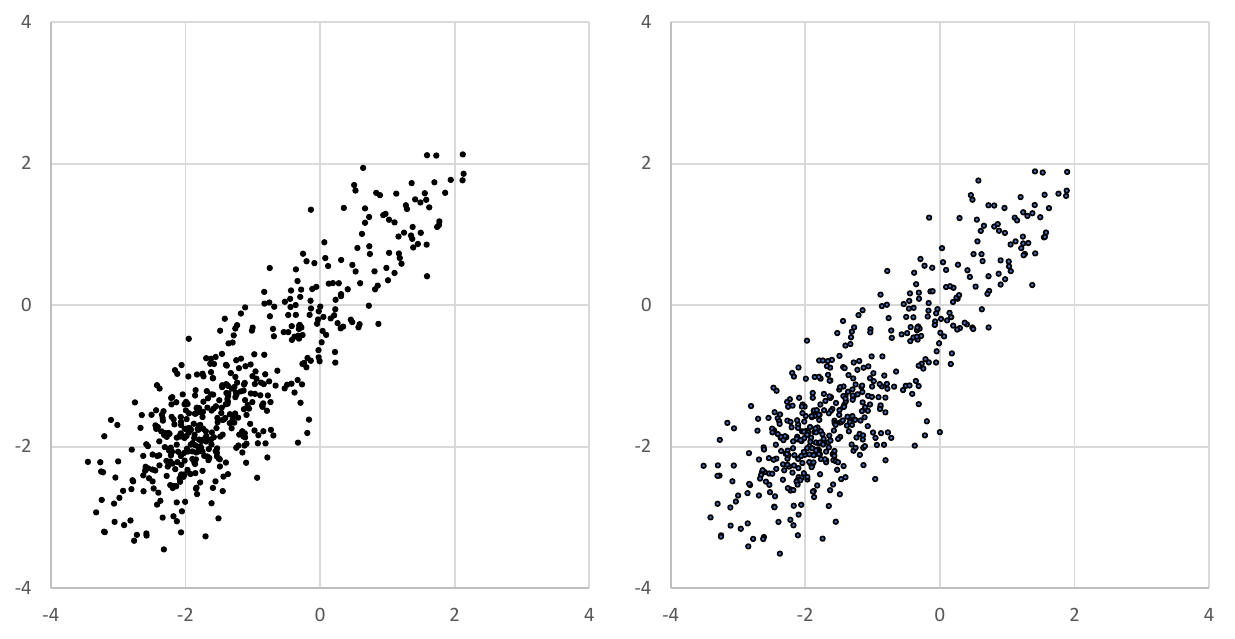}
\caption{Training data obtained from true model (left), AR model (center) and asymmetric AR model (right).}
\label{Fig_Training_data}
\end{center}
\end{figure}

In the following section, we present the results of estimating the latent signal $x_n$ from the observed time series $y_n$ using the GP-SSM. As training data for the Gaussian process, three cases are examined: sampled true latent signal, the signal estimated by the AR model, and the signal estimated by the asymmetric AR model shown in Figure \ref{Fig_Training_data}.

\begin{table}[tbp]
\caption{Estimated parameters of the Gaussian process state-space models}
\label{Tab_prameters_GP-SSM}
\begin{center}
\begin{tabular}{l|l|c|ccccc}
\makecell{training\\[-1mm]data} & kernel &log-likelihood & $\tau^2$ & $\sigma^2$ & $\theta_1$ & $\theta_2$ & $\theta_3$ \\
\hline
 true   & rbf        & -911.020 & 1.030 & 0.790 & 0.089 & ---   & 0.690 \\
 true   & linear+rbf & -911.614 & 1.239 & 0.238 & 0.439 & 0.0054& 0.656 \\
 true   & linear+rbf & -913.035 & 1.328 & 0.268 & 0.450 & 0.230 & 0.568 \\
 AR     & rbf        & -930.538 & 1.520 & 0.538 & 0.026 & ---   & 0.520 \\
 AR     & linear+rbf & -930.749 & 1.369 & 0.048 & 0.439 & 0.054 & 0.656 \\
 AS-AR  & rbf        & -928.464 & 1.506 & 0.557 & 0.044 & ---   & 0.492 \\
 AS-AR  & linear+rbf & -927.884 & 1.369 & 0.048 & 0.482 & 0.054 & 0.580 \\
\hline
\end{tabular}
\end{center}
\end{table}

Table \ref{Tab_prameters_GP-SSM} presents the estimation results obtained using GP-SSM with the rbf kernel and the linear + rbf kernel. The parameters were estimated by maximum likelihood. When the true training data were used, both the rbf kernel and the linear + rbf kernel achieved nearly the same level of fit as the particle filter results.

\begin{figure}[tbp]
\begin{center}
\includegraphics[width=30mm,angle=0,clip=]{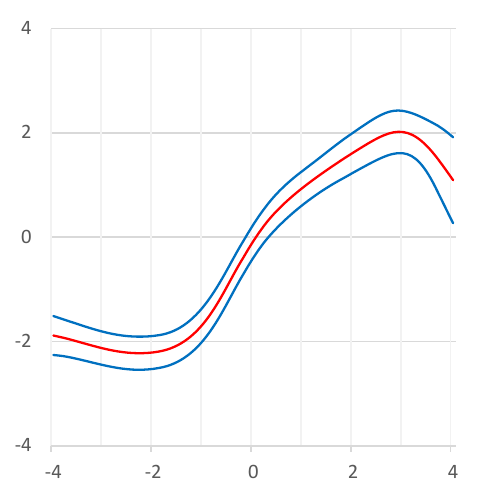}\hspace{5mm}
\includegraphics[width=100mm,angle=0,clip=]{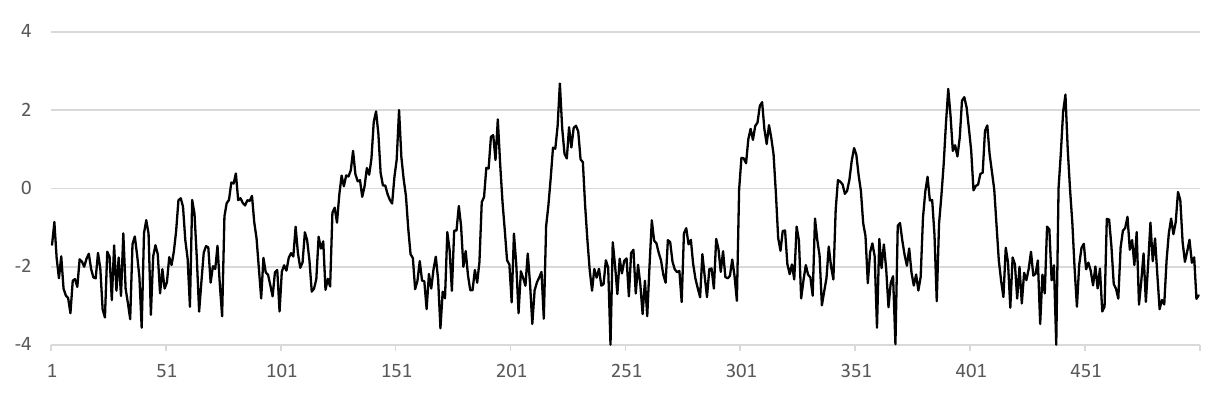}
\includegraphics[width=30mm,angle=0,clip=]{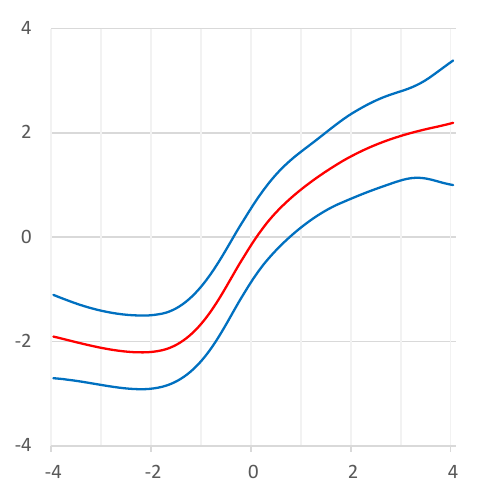}\hspace{5mm}
\includegraphics[width=100mm,angle=0,clip=]{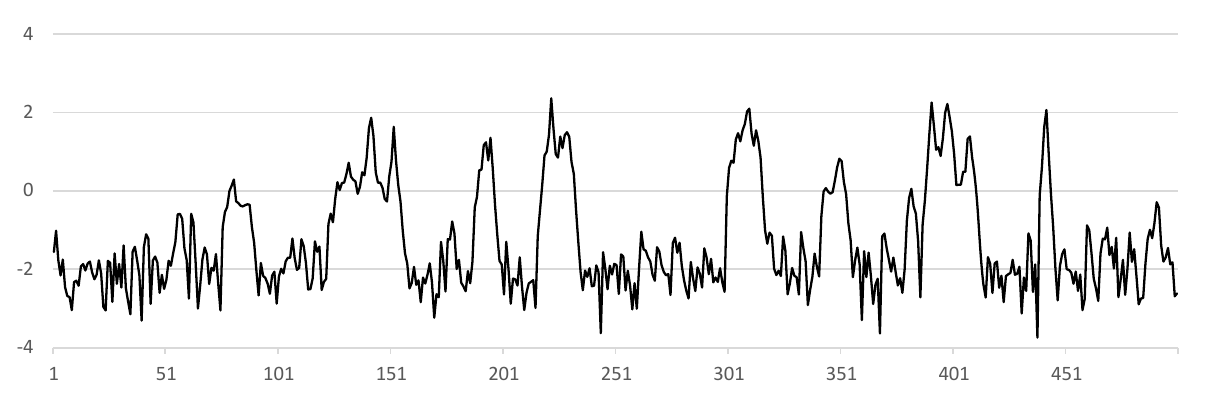}
\caption{GP and the signal $x_n$ estimated by the GP-SSM model based on the training data obtained from the true model and the estimated signal. Top plots: rbf kernel, bottom plots: linear-rbf kernel.}
\label{Fig_GP-SSM_true-kernel}
\end{center}
\end{figure}

Figure \ref{Fig_GP-SSM_true-kernel} shows the results of the GP-SSM using training data sampled from the true transition function $f(x)$.
Since the rbf kernel approaches zero in regions where no training data exist, the GP shown in the upper-left panel deviates substantially from the true function for $x>3$ , where the training samples are sparse.
In contrast, the GP-SSM using the linear-rbf kernel represents the underlying transition function $f(x)$ more accurately.
However, as indicated by the log-likelihood values, there is no significant difference between the estimated latent signals $x_n$ obtained using these kernels.

\begin{figure}[tbp]
\begin{center}
\includegraphics[width=30mm,angle=0,clip=]{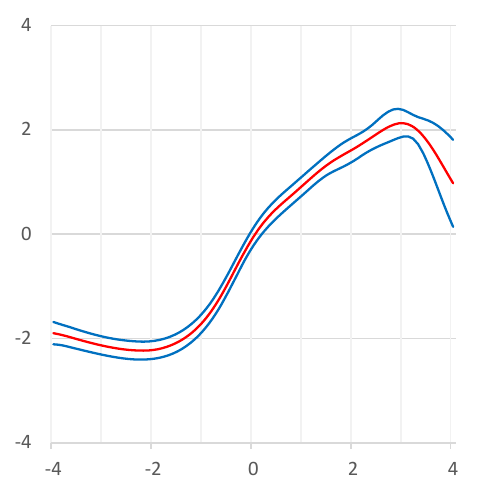}\hspace{5mm}
\includegraphics[width=100mm,angle=0,clip=]{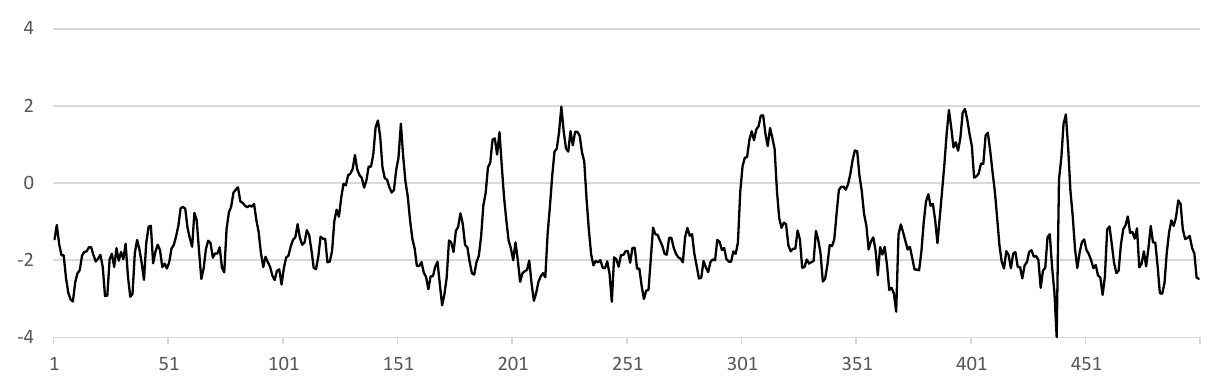}
\includegraphics[width=30mm,angle=0,clip=]{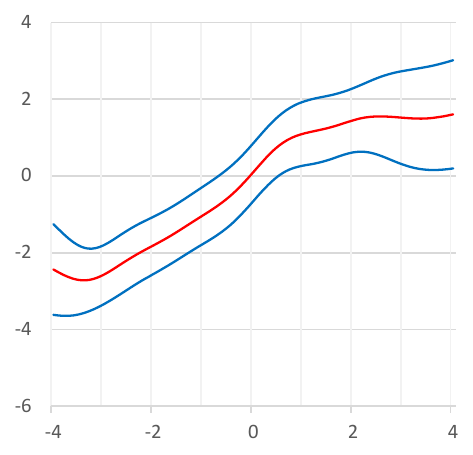}\hspace{5mm}
\includegraphics[width=100mm,angle=0,clip=]{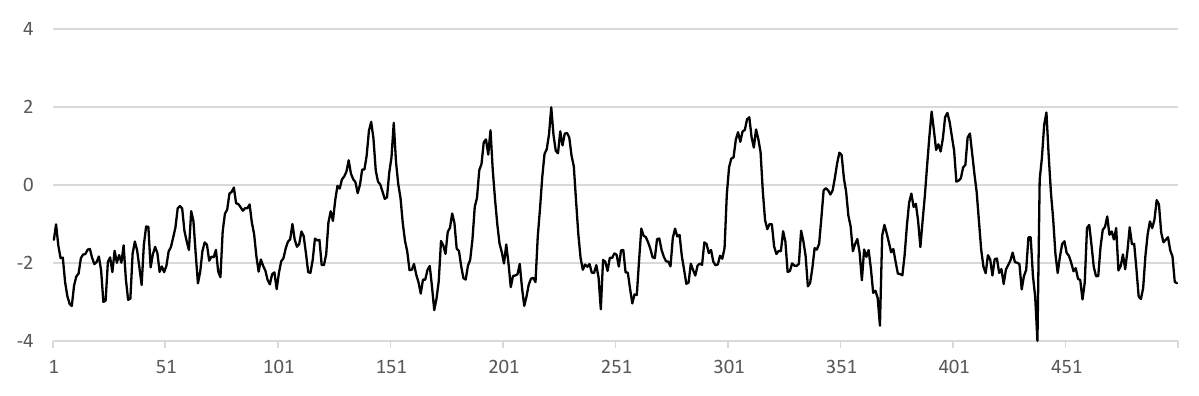}
\caption{GP and the signal $x_n$ estimated by the GP-SSM model based on the training data obtained from the AR model and the estimated signal. Top plots: rbf kernel, bottom plots: linear-rbf kernel.}
\label{Fig_GP-SSM_AR-training-data}
\end{center}
\end{figure}


\begin{figure}[tbp]
\begin{center}
\includegraphics[width=30mm,angle=0,clip=]{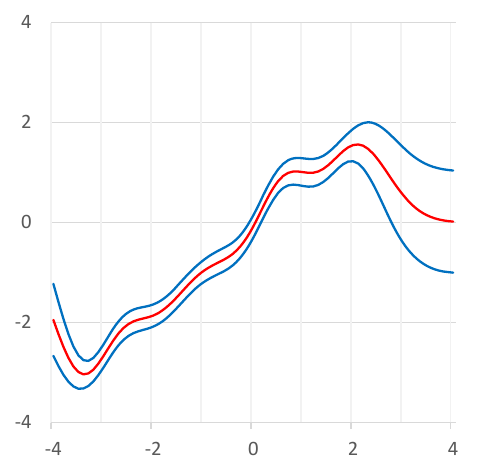}\hspace{5mm}
\includegraphics[width=100mm,angle=0,clip=]{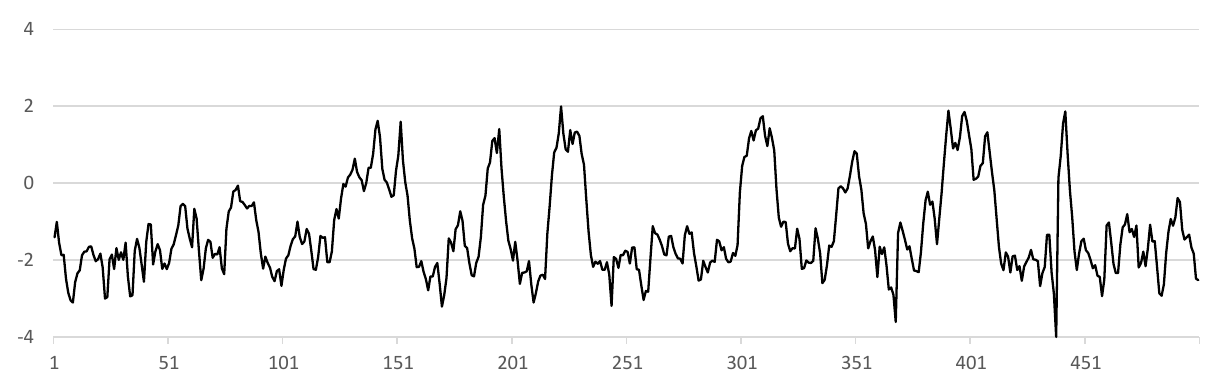}
\includegraphics[width=30mm,angle=0,clip=]{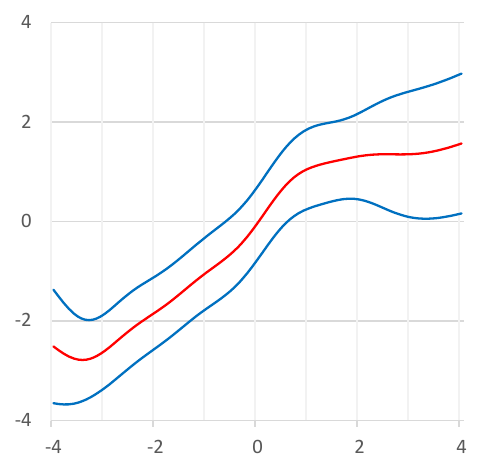}\hspace{5mm}
\includegraphics[width=100mm,angle=0,clip=]{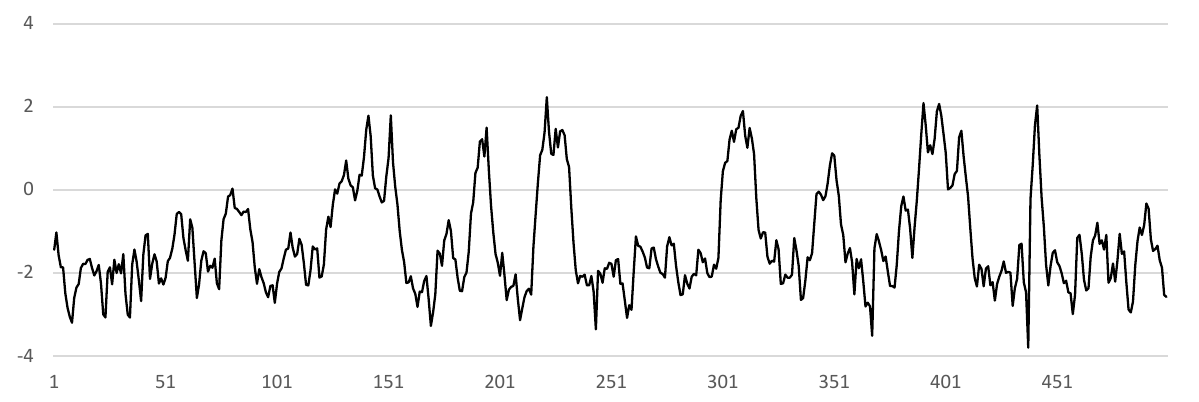}
\caption{GP and the signal $x_n$ estimated by the GP-SSM model based on the training data obtained from the asymmetric AR model and the estimated signal. Top plots: rbf kernel, bottom plots: linear-rbf kernel.}
\label{Fig_GP-SSM_asymmetric-AR-training-data}
\end{center}
\end{figure}

Figure \ref{Fig_GP-SSM_AR-training-data} presents the results of the GP-SSM when the signal estimated by the AR model is used as training data. The GP with the rbf kernel produces results similar to those obtained using the true training data. In the case of the linear-rbf kernel, the resulting GP is relatively similar to that based on the true training data; however, an upward bias is observed around $x=-2$. As in the case of the true kernel, there is no substantial difference between the estimates of $x_n$ obtained using these two kernels in the GP-SSM. Nevertheless, compared to the true-kernel case, the results exhibit less variability and yield smoother estimates.

Figure \ref{Fig_GP-SSM_asymmetric-AR-training-data} shows the results of the GP-SSM when the signal estimated by the asymmetric AR model was used as the training data.
The GP estimated with the rbf kernel exhibits a more complex shape than the case where the AR-based training data were used; however, no substantial differences are observed aside from this.

Based on these results, at least for this dataset considered in this subsection, the structure of the GP varies considerably depending on the training data and the choice of kernel. Nevertheless, the estimated state $x_n$ remains relatively stable across conditions.

\subsection{Nonlinear Smoothing}

\begin{figure}[tbp]
\begin{center}
\includegraphics[width=100mm,angle=0,clip=]{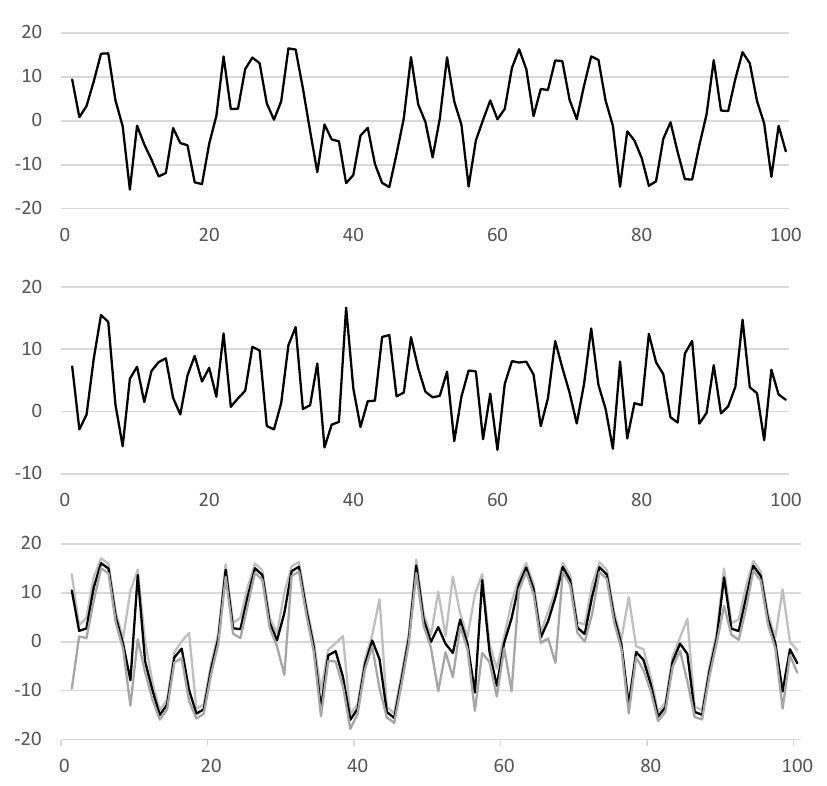}
\caption{Nonlinear data. Signal $x_n$, observed time series $y_n$ and the posterior distribution of the state $x_n$ obtained by particle filter.}\label{Fig_nlsim1-data}
\end{center}
\end{figure}

In this subsection, we consider the following nonlinear system with an external cosine input. Here, $x_n$ denotes the latent (unobserved) state, $y_n$  represents the observed time series, and $v_n$ and $w_n$ are white noise processes with $v_n \sim N(0,\tau^2)$ and $w_n \sim N(0,\sigma^2)$\cite{Kitagawa 1996}\cite{KG 1996}. 
The objective is to estimate the latent state $x_n$ from the observed sequence $y_n$,
\begin{eqnarray}
  \begin{array}{ccl}
    x_n & = & {\displaystyle\frac{1}{2}x_{n-1} + \frac{25x_{n-1}}{x_{n-1}^2 + 1} } + 8\cos( 1.2n ) + 
v_n \\
    y_n & = & \displaystyle\frac{x_n^2}{10} + w_n .
  \end{array}
\end{eqnarray}
The upper and middle panels of Figure \ref{Fig_nlsim1-data} present example realizations of $x_n$ and $y_n$, respectively, generated using random noise. For comparison with the GP-SSM, the lower panel shows the posterior distribution of the estimated latent states $x_n$, obtained using a particle filter under the assumption that the nonlinear model is known.

\begin{figure}[tbp]
\begin{center}
\includegraphics[width=100mm,angle=0,clip=]{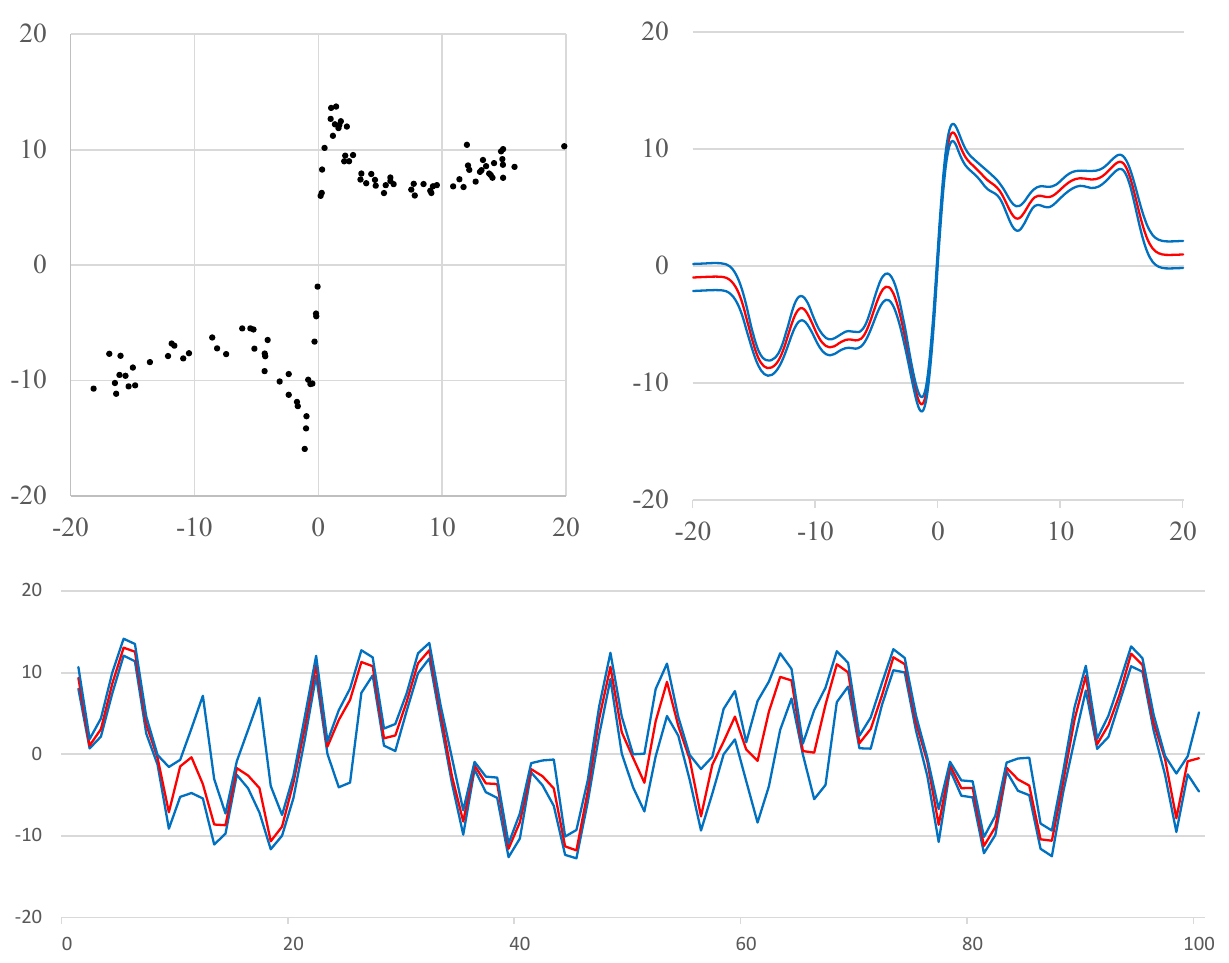}
\caption{Estimation of state by linear-rbf kernel}
\label{Fig_nlsim1_linear-rbf-kernel}
\end{center}
\end{figure}

Table \ref{Tab_prameters_GP-SSM_nlsim1} presents the results of fitting both the standard state-space model (SSM) and the GP-SSM.
In the SSM, the functional form and parameters are assumed to be known.
For the GP-SSM, two types of training data are used.
The data labeled ^^ ^^ true" correspond to noisy observations of the true transition function, as shown in the upper-left panel of Figure \ref{Fig_nlsim1_linear-rbf-kernel}.
The data labeled ^^ ^^ step function" are generated by adding noise to a step function with a jump at $x=0$.
In both cases, the number of training data points is 41.

\begin{table}[tbp]
\caption{Estimated parameters of the Gaussian process state-space models for nonlinear data}
\label{Tab_prameters_GP-SSM_nlsim1}
\begin{center}
\begin{tabular}{l|l|c|cccccc}
model & training data & kernel &log-likelihood & $\tau^2$ & $\sigma^2$ & $\theta_1$ & $\theta_2$ & $\theta_3$ \\
\hline
SSM    & ---             & ---        & $-281.043$ & 1.000    & 10.0 & --- & ---    \\
GP-SSM & true            & linear+rbf & $-313.820$ & 0.718E-3 & 17.1 & 0.104E-2 & 0.831E-2 & 9.100 \\
GP-SSM & true            & linear+exp & $-313.562$ & 0.500E-3 & 18.0 & 0.110E-2 & 0.210E-1 & 0.017 \\
GP-SSM & step function   & linear+rbf & $-313.491$ & 0.221E-3 & 16.8 & 0.070E-2 & 0.117E-1 & 8.500  \\
GP-SSM & step function   & linear+exp & $-313.413$ & 0.450E-3 & 17.6 & 0.082E-2 & 0.095E-1 & 0.027  \\
\hline
\end{tabular}
\end{center}
\end{table}

As kernel functions, two combined kernels are used: linear+rbf and linear+exponential (exp).
The reason for testing the exponential kernel instead of the rbf is that the exponential kernel can accommodate jumps more easily.

The log-likelihood of the GP-SSM is about 30 smaller than that of the SSM, which assumes the model is known.
The log-likelihoods of the linear+rbf and linear+exp kernels are almost identical.
Another interesting observation is that the results are nearly the same regardless of whether the training data are the ^^ ^^ true" set or the ^^ ^^ step-function" set.
This may be because what is essential in this system is the presence of a jump at the origin and the existence of stable points near $\pm 10$.

Upper left plot of Figure \ref{Fig_nlsim1_linear-rbf-kernel} shows the training dataset constructed as follows.
Using the posterior median estimates of $\hat{x}_n$ obtained in Figure \ref{Fig_nlsim1-data}, a two-dimensional paired dataset $(\hat{x}_{n-1}, \hat{x}_n-\cos (1.2n))$ was formed, from which every other sample was selected, yielding a total of 50 training data. The upper-right panel shows the resulting Gaussian Process with a linear-rbf kernel trained on this dataset.
Furthermore, the lower panel presents the posterior distribution of the estimated latent signal $x_n$ obtained by applying the GP-SSM based on this GP, where the mean and one standard deviation are displayed. The resulting posterior distribution is comparable to that obtained using the particle filter in Figure \ref{Fig_nlsim1-data}.

\begin{figure}[h]
\begin{center}
\includegraphics[width=100mm,angle=0,clip=]{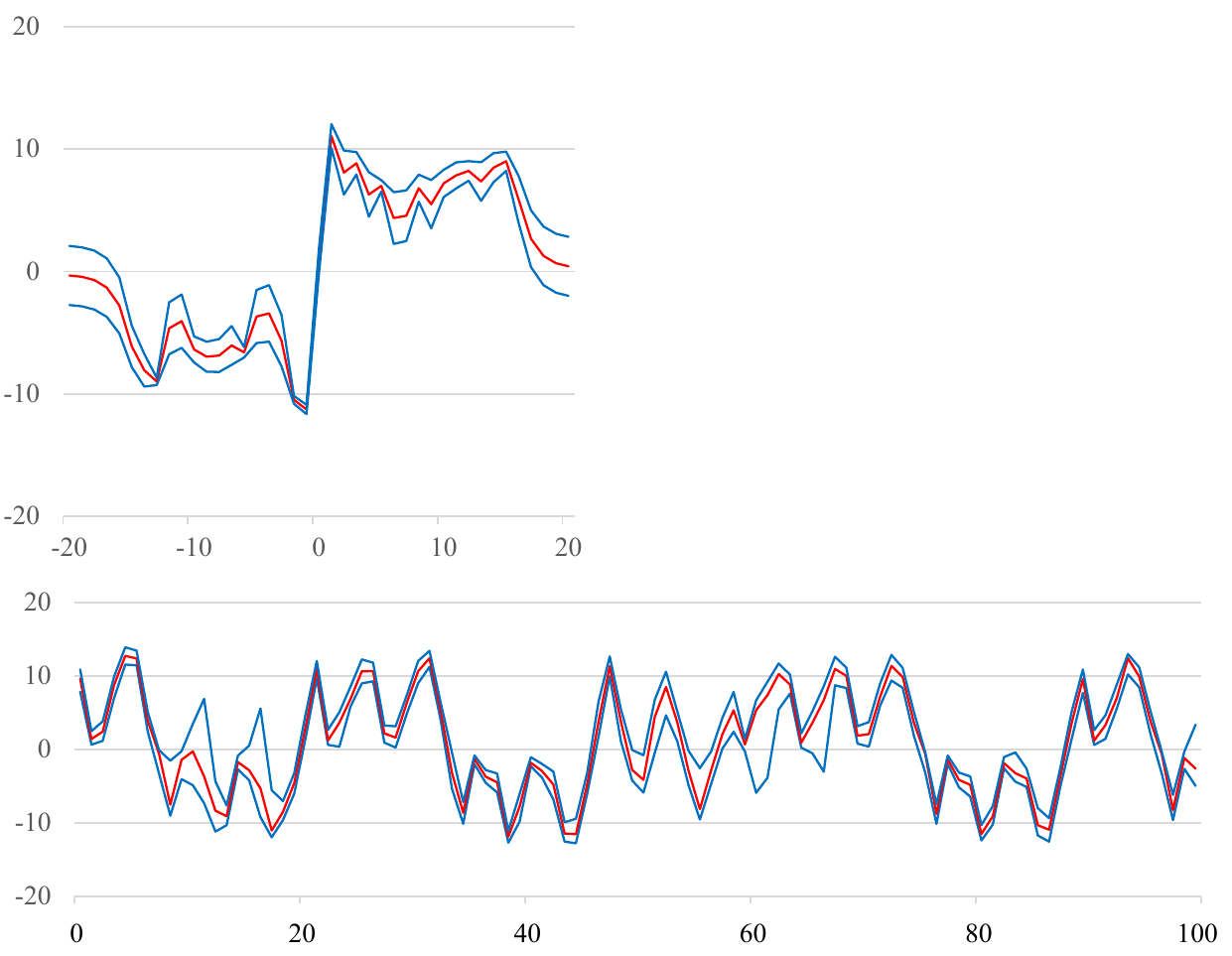}
\caption{Estimation of state by linear-exponential kernel}
\label{Fig_nlsim1_linear-exponential-kernel}
\includegraphics[width=100mm,angle=0,clip=]{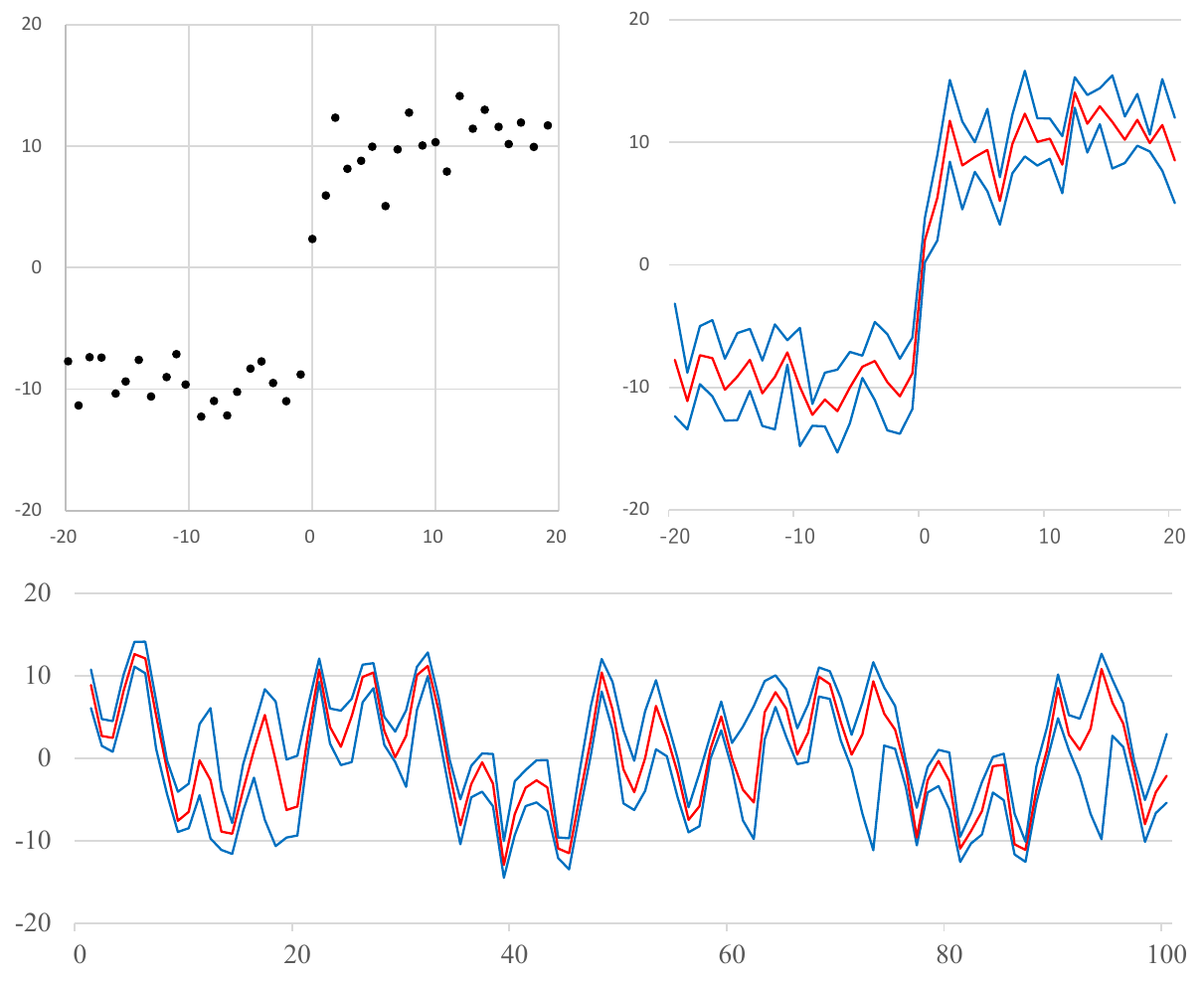}
\caption{Estimation of state by linear-exponential kernel. top-left plot: training data,
top-right: Gaussian process, bottom plot: posterior distribution of the signal $x_n$}
\label{Fig_nlsim1_test-training-data}
\end{center}
\end{figure}

Figure \ref{Fig_nlsim1_linear-exponential-kernel} presents the GP and the posterior distribution of the state $x_n$ obtained using a linear-exponential kernel trained with the same learning data. In this example, the nonlinear function $f(x)$ exhibits a sharp jump around the origin; therefore, the exponential kernel was used instead of the rbf kernel.

Up to this point, we assumed that the true model was known and generated the training data from the estimated latent state $x_n$ obtained by the particle filter. These training data were then used to construct the GP. However, such a situation is not realistic in actual applications. Therefore, we used an artificially generated step-shaped training dataset and estimated both the GP and the latent state $x_n$. The results are shown in Figure \ref{Fig_nlsim1_test-training-data}. Although the GP does not reproduce the sharp jump as clearly as in the previous figures, the posterior distribution of the latent state $x_n$ is reasonably close to those obtained in Figures \ref{Fig_nlsim1_linear-rbf-kernel} and \ref{Fig_nlsim1_linear-exponential-kernel}, despite having a slightly wider uncertainty range.

\section{Concluding Remarks}

By combining Gaussian process state-space models (GP-SSMs) with particle filtering, we can perform state estimation even when key components of a general state-space model such as the functional form of the nonlinear transition function and/or the noise variances are unknown. These quantities can instead be learned directly from data. Through four illustrative examples, we examined how well GP-SSMs perform compared with models whose structure is fully specified in advance.

In the examples on trend and seasonal components, we demonstrated that GP-SSMs can reproduce results comparable to those obtained from Kalman-filter-based linear models while relying on fewer structural assumptions, thereby offering a data-driven alternative when model specification is uncertain. In the example with asymmetric dynamics, we showed that GP-SSMs naturally capture regime-dependent behavior without requiring an explicit switching structure. Furthermore, in the system with strong nonlinearity and exogenous inputs, GP-SSMs successfully learned complex transition relationships that would be difficult to parametrize analytically.

Taken together, these findings confirm that GP-SSMs provide a unified and nonparametric approach to signal extraction and nonlinear filtering, particularly when the underlying transition mechanisms are unknown or highly nonlinear. Future research directions include improving computational efficiency, developing principled methods for model selection and training data construction, and extending the framework to higher-dimensional or hierarchical state-space models.


\begin{thebibliography}{2}

\bibitem{BMR 2016}
Bastani, V., Marcenaro, L. and Regazzoni, C.S., (2016), ^^ ^^ Incremental Nonlinear System Identification and Adaptive Particle Filtering Using Gaussian Process," arXiv:1608.08352v1.

\bibitem{DFG 2001}
Doucet, A.,  de Freitas, N., and Gordon, N., (2001).
    {\it Sequential Monte Carlo Methods in Practice}. 
    Springer-Verlag, New York.

\bibitem{DK 2012}
Durbin, J. and Koopman, S. J. (2012). \textit{Time Series Analysis by State Space Methods (2nd ed.)}. Oxford University Press.

\bibitem{ENDH 2017}
Eleftheriadis, S., Nicholson, T.F.W, Deisenroth, M.P. and Hensman, J. 2017, ^^ ^^ Identification of Gaussian Process State Space Models," \textit{31st Conference on Neural Information Processing Systems (NIPS 2017)}, Long Beach, CA, USA.

\bibitem{FLSR 2013}
Frigola, R., Lindsten, F., Scho\"{o}n, T. and Rasmussen, C.E. (2013), ^^ ^^ Bayesian Inference and Learning in Gaussian Process State-Space Models with Particle MCMC," arXiv: 1306.2861v2.

\bibitem{FCR 2014}
Frigola, R., Chen, Y. and Rasmussen, C.E. (2014), ^^ ^^ Variational Gaussian Process State-Space Models", 

\bibitem{GSS 1993}
Gordon, N.~J., Salmond, D.~J., and Smith, A.~F.~M., (1993). 
    Novel approach to nonlinear/non-Gaussian Bayesian state estimation, 
    {\it IEE Proceedings--F}, {\bf 140}, 107--113.

\bibitem{Kitagawa 1987}
Kitagawa, G. (1987). ^^ ^^ Non-Gaussian state-space modeling of nonstationary time series,"
{\it Journal of the American Statistical Association}, 82(400), 1032--1041.


\bibitem{Kitagawa 1996} 
Kitagawa, G. (1996). ^^ ^^ Monte Carlo filter and smoother for non-Gaussian nonlinear state space models," 
{\it Journal of Computational and Graphical Statistics}, \textbf{5}(1), 1--25.

\bibitem{Kitagawa 2021}
Kitagawa, G. (2021). \textit{Introduction to Time Series Modeling with Applications in R},
Second Edition, Chapman \& Hall CRC Press.

\bibitem{KG 1984} Kitagawa, G. and Gersch, W. (1984), ^^ ^^ A smoothness priors-state space modeling of
time series with trend and seasonality," {\it J. Amer. Statist. Assoc.}, {\bf 79}, 378--389.

\bibitem{KG 1996} Kitagawa, G. and Gersch, W. (1996), {\it Smoothness Priors Analysis of Time Series},
{\it Lecture Notes in Statistics}, {\bf 116}, Springer, New York.

\bibitem{Murphy 2012}
Murphy, K. P. (2012), \textit{Machine Learning: A Probabilistic Perspective}. MIT Press.

\bibitem{RW 2006}
Rasmussen, C. E. and Williams, C. K. I. (2006), \textit{Gaussian Processes for Machine Learning}. MIT Press.

\bibitem{SS 2017}
Shumway, R. H. and Stoffer, D. S. (2017). \textit{Time Series Analysis and Its Applications (4th ed.)}. Springer.

\bibitem{WFH 2017}
Wang, J.M., Fleet, D.J. and Hertzmann, A., (2017), ^^ ^^ Gaussian Process Dynamical Models,"
\end{thebibliography}
\end{document}